\documentclass{PoS}
\usepackage{url}
\usepackage{graphicx}
\title{Atmospheric calibration of the Cherenkov Telescope Array}

\ShortTitle{Atmospheric calibration of the CTA}

\author{\speaker{Jan Ebr}\\
Institute of Physics, Czech Academy of Sciences, Prague, Czech Republic\\
E-mail: \email{ebr@fzu.cz}}
\author{Tomasz Bulik\\
Warsaw University Astronomical Observatory, Warsaw, Poland.\\
E-mail: \email{tb@astrouw.edu.pl}}
\author{Llu{\'i}s Font\\
Unitat de F{\'i}sica de les Radiacions, Departament de Fiisica, and CERES-IEEC, Universitat Aut{\`o}noma de Barcelona, E-08193 Bellaterra, Spain.\\
E-mail: \email{lluis.font@uab.cat}}
\author{Markus Gaug\\
Unitat de F{\'i}sica de les Radiacions, Departament de Fiisica, and CERES-IEEC, Universitat Aut{\`o}noma de Barcelona, E-08193 Bellaterra, Spain\\
E-mail: \email{markus.gaug@uab.cat}}
\author{Petr Jane{\v c}ek\\
Institute of Physics, Czech Academy of Sciences, Prague, Czech Republic\\
E-mail: \email{janecekp@fzu.cz}}
\author{Jakub Jury{\v s}ek\\
Institute of Physics, Czech Academy of Sciences, Prague, Czech Republic\\
E-mail: \email{jurysek@fzu.cz}}
\author{Du{\v s}an Mand{\' a}t\\
Institute of Physics, Czech Academy of Sciences, Prague, Czech Republic\\
E-mail: \email{mandat@fzu.cz}}
\author{Stanislav {\v S}tef{\' a}nik\\
Faculty of Mathematics and Physics, Charles University, Prague, Czech Republic\\
E-mail: \email{stefanik@ipnp.troja.mff.cuni.cz}}
\author{Laura Valore\\
Universit{\`a} degli Studi di Napoli Federico II and INFN Napoli\\
E-mail: \email{valore@na.infn.it}}
\author{George Vasileiadis\\
LUPM, IN2P3/CNRS and Un.Montpellier II, France\\
E-mail: \email{Georges.Vasileiadis@umontpellier.fr}}
\author{for The CTA Consortium\\
\email{http://cta-observatory.org}}

\abstract{Atmospheric monitoring is an integral part of the design of the Cherenkov Telescope Array (CTA), as atmospheric conditions affect the observations by Imaging Atmospheric Cherenkov Telescopes (IACT) in multiple ways. The variable optical properties of the atmosphere are a major contribution to the systematic uncertainty in the determination of the energy and flux of the gamma photons. Both the development of the air-shower and the production of Cherenkov light depend on the molecular profile of the atmosphere. Additionally, the rapidly changing aerosol profile, affecting the transmission of the Cherenkov light, needs to be monitored on short time scales. Establishing a procedure to select targets based on current atmospheric conditions can increase the efficiency of the use of the observation time. The knowledge of atmospheric properties of the future CTA locations and their annual and short-term variations in advance is essential so that the atmospheric calibration can be readily applied to first scientific data. To this end, some devices are already installed at one or both of the selected sites. These include a Sun/Moon photometer, all-sky cameras for cloud detection and stellar photometry, detectors of night sky background and weather stations; a small optical telescope to measure atmospheric extinction using stellar photometry (FRAM) is ready to be deployed soon. Aerosol climatology at both sites will be soon assessed in a dedicated campaign using the ARCADE LIDAR. The atmospheric calibration strategy for the CTA contains dedicated instruments and methods for each task, namely: the combination of the FRAM and Raman Lidars to characterize the aerosol extinction profile along the line-of-sight of the observed target, the combination of an all-sky-camera and a ceilometer to characterize the atmosphere in the direction of possible future observations. These will be complemented by data from satellites and global data assimilation models (validated by a radiosonde campaign). Moreover, information about the atmosphere can be extracted directly from the data using the Cherenkov Transparency Coefficient (CTC) method.}

\FullConference{35th International Cosmic Ray Conference - ICRC2017\\
10-20 July, 2017\\
Bexco, Busan, Korea}

\begin{document}

\section{Introduction}

The Cherenkov Telescope Array (CTA) will be the next-generation Very High Energy (VHE) gamma-ray observatory for the observation of gamma-rays ranging from few tens of GeV to several hundreds of TeV. It stands in the tradition and improves the previous generation of Imaging Atmospheric Cherenkov Technique (IACT) experiments H.E.S.S., MAGIC and VERITAS. The technique being ground-based, permits the use of large mirrors and hence takes advantage of the huge areas illuminated by gamma-ray induced air showers, through the emission of Cherenkov light by the ultra-relativistic shower particles. It is therefore possible achieve unprecedented sensitivities in that energy range, orders of magnitude better than what can be achieved by satellites~\cite{cta}. 

Exploiting the atmosphere as a calorimeter comes however along with a series of limitations, on the accuracy of the energy and exposure reconstruction, as well as the precise absolute pointing. Current IACTs achieve around 15\% for the accuracy with which the energy scale is calibrated~\cite{magicperformance2,hessperformance}. These estimates have been obtained, however, with selected data and should be considered rather lower limits for the more general case of data taken under non-optimal, but still acceptable conditions. Moreover, current IACTs lose a considerable amount of their data during offline data selection, using parameters which correlate with the presence of clouds and aerosols. Finally, all these calibration difficulties increase with the zenith angle under which the target field is observed. 

Several atmospheric parameters affect the systematic uncertainties of IACTs~\cite{bernloehr}: the molecular density profile modulates the air-shower development along its trajectory and the Cherenkov angle, leading to non-linear and energy-dependent effects~\cite{bernloehr,vrastil}. Next, Cherenkov photons are either absorbed by molecules on their way towards the telescopes or scattered out of the camera field-of-view (FOV) by molecules or aerosols and clouds. The latter are typically thinner than the length of air showers affecting only a part of the emitted Cherenkov light. The altitude of a cloud can hence cause energy dependent effects~\cite{garridoicrc,nolan2010} (since the mean altitude of the shower maximum depends on the energy of the incident gamma-ray) and must be assessed, e.g. with a LIDAR~\cite{fruckatmohead}. 

A detailed atmospheric calibration strategy for the CTA has been presented recently~(see~\cite{gaug} and references therein). While previous IACTs and the Pierre Auger Observatory have pioneered several ways to introduce atmospheric calibration devices~\cite{atmomagic,atmohess,atmoauger,atmoveritas}, the CTA Central Calibration Facilities work-package intended to learn as much as possible from the experience of these and establish a coherent atmospheric calibration strategy. One of the key elements of this strategy consists in assessing the main characteristics of each site well in advance and build then tailored devices and methods to monitor each of these parameters. Contrary to previous IACTs, these must cover a wide field-of-view, of more than 10$^\circ\times 10^\circ$, and assess the extinction profile to a precision of a few percent. Moreover, we have learned that the scheduling strategies of current IACTs have still a great potential of improvement, if atmospheric conditions were taken into account for the specific science case behind an observation. This applies particularly to sources with very soft spectra observable below about 100~GeV only, and which require absolutely clear nights (so-called ``photometric nights'' in optical astronomy) in order to escape later offline data rejection. Similar arguments apply to precision pointing observations. While data from other targets at higher energies can be rather easily corrected, these low-energy events are lost since they do not even trigger the readout. 

\begin{figure} 
\centering
\includegraphics[width=.5\textwidth]{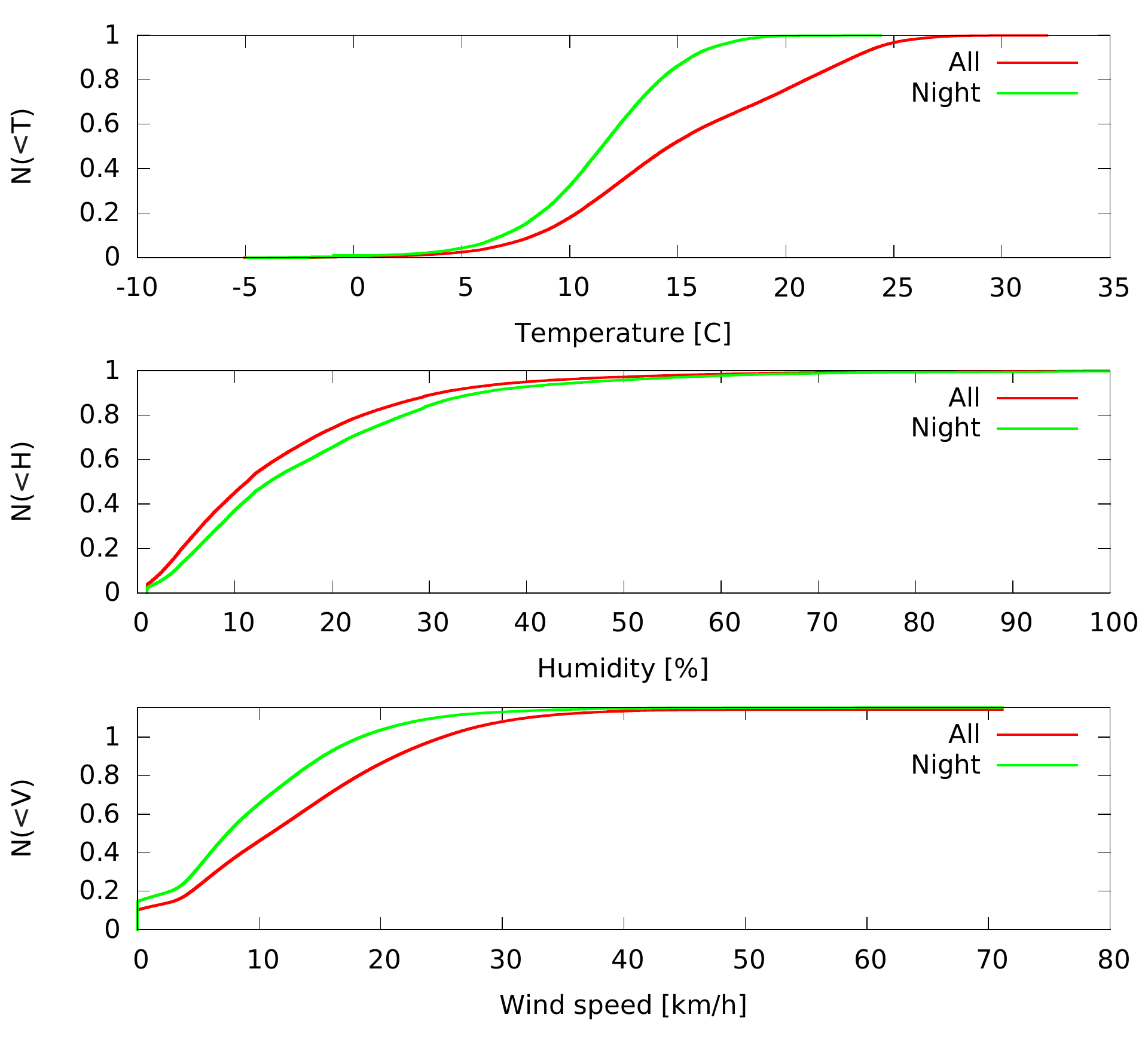}
\caption{Cumulative distributions of temperature, humidity and wind at 10 meters
observed at the southern CTA site.
For each quantity we present separately the night time distribution and
the one that encompasses all values measured - day and night}
\end{figure}

Currently, various on-site characterization activities are carried out to determine the atmospheric properties of both sites. These are needed to establish specifications for the atmospheric monitoring systems and to take few remaining design choices. The sites will be characterized with the help of a radio sonde campaign to assess molecular profiles, and particularly the accuracy of global data assimilation systems predicting the density profiles, moreover with a dedicated LIDAR project to assess aerosol profiles~\cite{arcade}, and a Sun/Moon photometer~\cite{photometer}. Dedicated systems, which will later belong to the CTA observatory, used to continuously characterize the observed science field-of-view, contain a Raman LIDAR~\cite{vasileiadis} and a small optical telescope to measure atmospheric extinction using stellar photometry (the co-called FRAM). Instruments assisting the scheduler contain an All-Sky Camera (ASC)~\cite{adam} and a commercial Ceilometer to assess cloud coverage and their altitudes, respectively. Some of those systems are also being gradually deployed on the sites to assist in their characterization.

\section{Site characterization}

Currently, at the southern site in Chile, there is a {\bf site characterization station}. The station includes the following instruments: a 10-meter mast with a Reinhardt weather station, a 30-meter tower with three 3-dimensional anemometers, a seismometer, an All-sky camera (ASC), web-cams, and a Sun/Moon photometer. The initial installation started in February 2015 with the 10-meter mast and a weather station. This was followed with the installation of the remaining instruments a year later. The weather station on the 10-meter mast records temperature, humidity, wind, and air pressure continuously every minute, but we also keep record of the wind every two seconds. The 3-dimensional anemometers on the 30-meter tower record the wind every second, and are located at 10, 20 and 30 meters above the ground. The seismometer is installed on a concrete platform which is not attached to the bedrock in order to simulate the seismic effects on the telescope build on floating foundations. 

The {\bf All Sky Camera} (ASC) is a passive instrument for the night-time detection of clouds and a coarse atmospheric extinction study of the CTA sites. The camera consists of an astronomical CCD camera G2-4000 with Johnson BVR filters and a cooled CCD chip. The resolution of the CCD chip of the camera is $2056\times2062$ pixels, the angular resolution is 0.1 degrees and the lens $f$-number is 2.8 (Sigma 4.5/2.8 EX DC). We observe stars up to the 7th magnitude in zenith with 30 seconds of exposure time typically. Both CTA sites are equipped with the same type of ASC and an extensive site characterization study of clouds, their behavior and general night sky photometry is ongoing from 10/2015 (La Palma), 11/2015 (Chile) respectively; each ASC takes an image of the sky every five minutes each night \cite{dusan}. The main task of the instrument during the CTA operation will be the analysis of the clouds in the sky above the CTA observatories before and during each observing time slot. The goal is to monitor the cloud conditions and predict their short-term changes. The photometric BVR Johnson filters installed allow us to measure the flux from the known stars and compare the signal with the catalog. The TYCHO2 star catalog is used both for the astrometry and clouds analysis and for star photometry.

The {\bf Sun/Moon photometer CE318-T} is already installed at the Southern CTA site for long-term monitoring of atmospheric conditions, characterization of performance of the instrument and testing methods of nocturnal 
AOD (Atmoshperic Optical Depth) retrieval --  it has been taking data in 2016, and then was temporarily taken out of the site for calibration. The photometer is part of AERONET (AErosol RObotic NETwork) of precisely calibrated and globally distributed instruments, providing precise measurements of diurnal AODs. This type of photometer can measure diurnal and nocturnal evolution of the integral optical depth of the atmosphere with very high precision in 9 photometric pass-bands with high cadence of one measurement every 3 minutes. Using the Sun or the Moon as sources of light, the photometer overcomes the usual disadvantage of low signal-to-noise ratio of measurements of passive sensors pointing on stars. Based on the first analysis of two months of data from the Southern CTA site, uncertainties of nocturnal AODs measured by the photometer range between 0.008--0.015 in the 500 nm pass-band. 
Although the calibration and AOD retrieval from the Sun photometry is well understood, several problems persist in the case of Lunar photometry. Since the irradiation of the Moon varies with time and the photometer is not temperature stabilized, the calibration of the photometer on site is much more difficult. The photometer for CTA, however, is installed near the Cerro Paranal, a place with very stable atmospheric conditions at high altitude and its data can therefore be used for testing new methods of AOD retrieval, calibration and cloud-screening. Substantial progress has been made recently and a first complete analysis of diurnal and nocturnal AODs for two months of observations from the Southern CTA site is now available. For a detailed description of methods and obtained results see the accompanying contribution \cite{photometer}.

The {\bf ARCADE} system is a Raman Lidar that will characterize the two CTA sites (first in La Palma for about one year, then in Chile), before and during the construction of the array, to collect data on the aerosol properties on site. The measurements of the aerosol stratification will help to define Monte Carlo simulations of the shower development in atmosphere under different atmospheric conditions and the detector response. ARCADE will be also to directly cross-calibrate the future Raman Lidars that are expected to operate at the CTA sites during data taking. The system has been recently upgraded and is presently in L'Aquila (Italy) for its validation before shipping to La Palma. For more details, see the accompanying contribution \cite{arcade}.

The {\bf FRAM} is a small optical telescope (or rather a 135mm photographic lens attached to a large $36\times36$ mm CCD camera) which will be used during the operation of the Observatory to assess the immediate conditions in the field(s) of view and their changes on timescales of minutes, using stellar photometry to provide maps of atmospheric extinction due to aerosols and clouds across the current field(s) of view of the IACTs. In regular intervals, series of images across different altitudes (scans) will be taken for self-calibration of the instrument, providing incidentally high-precision measurements of the integral aerosol optical depth. For site characterization, the FRAMs will be deployed in advance of the observatory construction at both sites and perform these scans continuously throughout the nights to investigate what temporal and spatial changes in the aerosol conditions are typical for the sites and to improve the development of analysis methods tailored to the conditions on the sites. For the southern CTA site, the hardware has been already shipped from Europe (in 05/2017), for the northern site, the formalities for deployment are being finalized (as of 06/2017) \cite{FRAM}.

\section{Further devices and methods in development}

A site-specific {\bf weather monitoring, modeling and forecasting system}, optimized for CTA, will be deployed to ensure the safety of humans, instruments and data collected. Such a system requires absolutely reliable functioning, particularly during periods of adverse or even dangerous weather conditions. For this reason, sufficient redundancy in the system, and intelligent detection of malfunctioning is a must. Three 10-m weather towers surrounding the central area reserved for the LSTs (Large Size Telescopes), will be constructed and equipped with weather monitoring instruments. Additionally, another circle, equipped with another three anemometers will measure the winds affecting the outer circle of MSTs (Medium Size Telescopes) and SSTs (Small Size Telescopes). As noted above, the wind height profile is currently assessed using various anemometers located at different heights on the 30-m tower, in order to reliably predict maximum wind speeds expected at altitude of the LST cameras, if measured at 10~m altitude. Additionally, lightning sensors and dust counters will be installed to assist the online detection of abnormal, or potentially dangerous situations. All equipment will be included in the fibre-optics communication system, in the alert system and the Graphical User Interface (GUI). The CTA will be also equipped with a direct link to modern data assimilation models, like the GDAS\footnote{https://www.ncdc.noaa.gov/data-access/model-data/model-datasets/ global-data-assimilation-system-gdas} or the ECMWF\footnote{http://www.ecmwf.int} and incorporate their now-casts of atmospheric density profile prediction into the data analysis pipeline.

For the last four years, the IFAE/UAB (Spain), the LUPM (France) and the CEILAP (Argentina) teams have jointly developped a {\bf Raman Lidar} project.
The goal is to build three identical Raman type Lidars, two to be installed in the Southern site, and one in the Northern CTA site. They will be operated in Raman mode,
permitting this way to reduce the systematic uncertainties of the aerosol profile of the atmosphere to less than 2\% at the moment of data taking,
which ensures, together with the overall extinction follow-up by the FRAM, that the CTA can meet the rather strict requirement of 10\% on the uncertainty with
which the absolute energy scale is known.

The European Lidars are based on refurbished containers and the telescopes from the CLUE experiment. They use a UV sensitive 1.8~m mirror and powerful Nd:YAG lasers as common elements. The Argentinian Lidar instead is based on a simultaneous use of several smaller mirrors focusing the backscattered light into standard fibres, from which the signals are then added optically. These projects are currently completing their prototypes, or have already started long-term testing of these, e.g. the Argentinean Lidar which has been brought to the Pierre Auger Observatory for such cross-calibrations and tests. Preliminary Monte-Carlo simulations and a detailed link-budget study carried out by the Spanish groups, have shown the feasibility of the European project. Later this year, long-term tests will take place on La Palma, to finalize the solution to be adopted for the contribution to CTA. For more details, see the accompanying contribution \cite{vasileiadis}.

The height of the clouds detected by the ASC can be assessed using a {\bf Ceilometer} -- a commercially available infrared Lidar, which can be used independently of the CTA telescopes as it does not interfere with their operation. The device considered is Vaisala CL51, capable of detection of three separate cloud layers up to 13 kilometers from the device. As the Ceilometer is usually used in a fixed vertical position (for example to measure the cloud base near an airport), a tailored weather-proof altitude-azimuth mount has been designed to allow directional measurements. 

Another method that will be used to monitor the atmospheric conditions once the Observatory is operational is based on the {\bf Cherenkov Transparency Coefficient} (CTC), which was successfully implemented in the H.E.S.S. experiment as a data quality criterion assessing the extinction of the Cherenkov light in the atmosphere~\cite{Hahn}. Using only the data gathered during scientific observations, the method is based on the monitoring of the system trigger rate which is sensitive to the immediate atmospheric conditions over the array.
Eliminating the contribution of the system to the stereo trigger rate, the robustness of the CTC is ensured under both short and long-term changes in the atmospheric transparency.
Maintaining the independence from hardware and observation related effects is a challenging task in CTA due to the diversity of installed systems, such as different optical systems, detector hardware, sizes of telescopes and, specifically, more complex array layouts.
Other influence that has not been addressed before is the possibility to change trigger thresholds during the data taking which is foreseen for one of the telescope types in CTA.
Currently, a Monte Carlo study of the CTC performance in CTA under different atmospheric and hardware conditions is under development.
Since the CTC will utilize the output of regular data taking in CTA, it will allow a fast evaluation of the atmospheric conditions and monitoring of the aerosol concentrations.
In order to ensure the consistency of the data quality selection and the hardware independence, the CTC values of the atmospheric transparency will be used to cross-check with the reports by specific atmospheric monitoring devices deployed in both CTA sites (e.g. Lidars, FRAM and AERONET Sun/Moon photometer). The preliminary results of a feasibility study for the application of the CTC in CTA are reported in a separate contribution in these proceedings~\cite{Stefanik}.

\section*{Acknowledgements}

 We gratefully acknowledge financial support from the agencies and
organizations listed here: \href{http://www.cta-observatory.org/consortium\_acknowledgments}{www.cta-observatory.org/consortium\_acknowledgments}, in particular by MEYS of the Czech Republic under the projects LTT17006 and LM2015046 and by ESIF and MEYS under the project CZ.02.1.01/0.0/0.0/16\_013/0001403 and  MINECO National R+D+I through the project FPA2015-69210-C6-6-R, FPA. This work was conducted in the context of the CTA Consortium.

\end{document}